\documentclass[a4paper,fleqn,usenatbib]{mnras}

\usepackage{mathptmx}

\usepackage[T1]{fontenc}
\usepackage{ae,aecompl}

\usepackage{graphicx}	
\usepackage{amsmath}	
\usepackage{amssymb}	
\usepackage{cite}
\usepackage{caption}

\title[GRO J1744-28: the Slowest Transitional Pulsar?]{The Bursting Pulsar GRO J1744-28: the Slowest Transitional Pulsar?}

\author[J. M. C. Court et al.]{J.M.C. Court$^{1}$, D. Altamirano$^{1}$ \& A. Sanna$^{2}$
\\
$^{1}$Department of Physics and Astronomy, University of Southampton, Southampton, SO17 1BJ, UK\\
$^{2}$Dipartimento di Fisica, Universit\`{a} degli Studi di Cagliari, SP Monserrato-Sestu km 0.7, 09042 Monserrato, Italy\\
}

\begin{document}
\maketitle

\begin{abstract}
GRO J1744-28 (the Bursting Pulsar) is a neutron star LMXB which shows highly structured X-ray variability near the end of its X-ray outbursts. In this letter we show that this variability is analogous to that seen in Transitional Millisecond Pulsars such as PSR J1023+0038: `missing link' systems consisting of a pulsar nearing the end of its recycling phase. As such, we show that the Bursting Pulsar may also be associated with this class of objects. We discuss the implications of this scenario; in particular, we discuss the fact that the Bursting Pulsar has a significantly higher spin period and magnetic field than any other known Transitional Pulsar. If the Bursting Pulsar is indeed transitional, then this source opens a new window of oppurtunity to test our understanding of these systems in an entirely unexplored physical regime.
\end{abstract}




\section{Introduction}

\par Millisecond Pulsars are old radio pulsars with spin periods of order $\sim10$\,ms \citep{Backer_MSP}. They have long been believed to be the end product of systems containing a neutron star (NS) in a Low Mass X-ray Binary (LMXB). In these systems, matter from a Roche-lobe overflowing star donates angular momentum to a NS, spinning it up to frequencies of several 100 Hz \citep{Alpar_MSP}. A number of fast-spinning X-ray pulsars (accreting Millisecond Pulsars, or AMXPs) have been found in LMXBs (e.g. \citealp{Wijnands_XRPulsar,Altamirano_Broken,Patruno_AllAMXPs,Sanna_AMXP}), seemingly confirming this physical picture. At the end of this so-called `recycling' process, the system should transition from an accretion-powered pulsar to a rotation-powered pulsar. As such, it has long been expected that such a transition could be observed by finding a system which changes its character from an accreting NS at one time to a radio pulsar at some later time. Subsequently a small family of 7 candidate objects have been discovered or proposed: these are referred to as Transitional Millisecond Pulsars (TMSPs).
\par The first of these objects, \textbf{PSR J1023+0038}, was identified by \citealp{Archibald_Link}. Although it appeared as a non-accreting radio pulsar at the time of identification in 2009, previous optical studies showed that this system contained an accretion disk in 2002 \citep{Szkody_1023Accretion}. As such, the pulsar in this system must have switched from an accreting phase to a radio pulsar phase at some point between 2003 and 2009, strongly suggesting the identification of this system as a TMSP. The pulsar in this system has a spin period of 1.69\,ms, and the companion is a star with a mass between $\sim$0.14--0.42\,M$_\odot$. \citealp{Archibald_Link} suggested that the low X-ray luminosity of PSR J1023+0038 in its accreting phase was due to accretion taking place in the `propeller regime' \citep{Illarionov_Propellor}. In this regime, accreting matter is halted by magnetic pressure above the co-rotation radius of the NS magnetosphere. This matter is then ejected from the system as a wind. Whether a system is in the propeller regime depends on its spin and its magnetic field strength \citep{Lewin_QPORev}. Additionally, below a certain accretion rate, no stable balance between ram pressure and radiation pressure can form and any disk is ejected from the system (e.g. \citealp{Campana_NoDisk}). \citealp{Archibald_Link} suggested that the current accretion rate in PSR J1023+0038 is only slightly below this critical value, and that any small increase in accretion rate could cause accretion in this system to resume. They suggested the possibility of TMSP systems which flip back and forth between accreting and radio pulsar phases multiple times.
\par \citealp{Papitto_Swings} identified \textbf{IGR J18245-2452} as the first pulsar to switch from a radio pulsar to an AMXP and back to a radio pulsar.  This source was first observed as a radio pulsar \citep{Manchester_PulsarCat}, before being observed several years later by \textit{XMM-Newton} \citep{Eckert_IGRJ18245} as an AMXP. Several months after the \textit{XMM-Newton} observation, \citealp{Papitto_Finding} found that the source had reactivated as a radio pulsar during X-ray quiescence. The pulsar in this system has a period of 3.93\,ms, and the companion star has a mass of $>0.17$\,M$_\odot$ \citep{Papitto_Swings}. During the 2013 outburst of IGR J18245-2452, \citealp{Ferrigno_TMSPVar} reported the presence of high-amplitude variability in the X-ray lightcurve. They interpreted this as being due to the accretion rate $\dot{M}$ being very close to the critical rate at which the propeller effect begins to dominate the flow geometry. In this regime, small fluctuations in $\dot{M}$ cause so-called `hiccups', in which matter alternates between being ejected by the propeller effect and being accreted onto the NS poles. Similar X-ray variability has subsequently been found in lightcurves from outbursts during the accreting phase of PSR J1023+0038 \citep{Bogdanov_TMSPVar}, suggesting that this variability is somehow intrinsic to TMSPs as a class of objects.
\par \textbf{1FGL J1227.9-4852} was first identified in the first \textit{Fermi}/LAT source catalogue \citep{Abdo_Catalogue}. \citealp{Hill_XSS} found that the $\gamma$-ray spectral characteristics of this source are consistent with known millisecond radio pulsars, although no radio pulsations were found. They suggested that this object could be associated with the X-ray source XSS J12270-4859. Before 2009, XSS J12270-4859 showed optical emission lines typical of an accretion disk \citep{Pretorius_Optical}. \citealp{Hill_XSS} suggested that XSS J12270-4859 may also be a TMSP, which switched from an accreting phase to a radio pulsar millisecond pulsar phase between 2009 and 2011. Subsequent studies have found pulsations in both the radio \citep{Roy_12270Spin} and $\gamma$-ray \citep{Johnson_12270Spin} emissions of this source, confirming the system contains a pulsar and establishing its spin period at 1.69\,ms.
\par \textbf{XMM J174457-2850.3} is a neutron star X-ray binary. Although no X-ray or radio pulsations have been detected due to the faintness of the source, \citealp{Degenaar_174457} have found that the X-ray variability properties of this source are similar to those seen in other TMSPs. This object also exhibits extended low-luminosity states during outbursts, which \citealp{Degenaar_174457} suggest may be symptomatic of TMSPs.
\par \textbf{3FGL J1544.6-1125} was also first identified in \textit{Fermi}/LAT data. \citealp{Bogdanov_Proxy} associated this object with the X-Ray source 1RXS J154439.4-112820. Due to the presence of $\gamma$-rays, as well as the presence of variability in the X-ray lightcurve similar to IGR J18245-2452, they proposed that this object is a TMSP in the accreting state. However, no pulsations from this system have been detected in the X-ray or the radio, so the pulsar period is not known. \citealp{Bogdanov_Proxy} found a bimodality in count rate during the period of X-ray variability, suggesting that this behaviour can be explained as quick transitions between three quasi-stable accretion modes known as `low' , `high' and `flaring'. This effect has also been seen in the TMSP IGR J18245-2452 \citep{Ferrigno_TMSPVar}.
\par \citealp{Strader_6} identified the $\gamma$-ray source, \textbf{3FGL J0427.9-6704}, as a TMSP. They found that this source also displays X-ray variability similar to what is seen from the other known TMSPs. Finally, \citealp{Rea_J0838} have proposed that the X-ray source \textbf{XMM J083850.4-282759} may also be a TMSP. Although this source has not been detected in the gamma or the radio, the authors argued that X-ray variability coupled with X-ray flaring seen from this object is reminiscent of similar behaviour seen in other TMSPs during subluminous disk states.
\par The phenomenology of currently known TMSPs is varied, and different methods have been used to conclude (or propose) that each individual system belongs to this class. The fact that 6 of the 7 objects show similar patterns of X-ray variability during outburst suggests that this variability can be used as an indication that a system may be a TMSP. In this letter we present evidence that a 7$^\mathrm{th}$ object, the so-called `Bursting Pulsar' \textbf{GRO J1744-28}, may also be associated with this family of objects.

\vspace{-2em} 

\section{GRO J1744-28: The Bursting Pulsar} \label{sec:BP}

\par GRO J1744-28, or the `Bursting Pulsar' \citep[hereafter BP,][]{Paciesas_BPDiscovery}, is an LMXB containing a 2.4\,Hz pulsar and a $\sim0.07-0.2$\,M$_\odot$ companion star \citep[e.g.][]{Sturner_BPNature,Finger_BP,Rappaport_BPHistory,Degenaar_BPSpec}; the orbital period of the systems is 11.8\,d \citep{Finger_Pulse,Sanna_BP}. Estimates for the NS magnetic field range between 2 and $50\times10^{10}$\,G \citep{Finger_BP,Degenaar_BPSpec,Dai_OB3,Doroshenko_NBFlash}. The BP is one of only two known systems that show `Type II' X-ray bursts: bright, $\sim$10\,s-long non-thermonuclear X-ray flares. The ``Rapid Burster'' (i.e. MXB 1730-33; hereafter RB), is the other system (e.g. \citealp{Lewin_TypeII,Kouveliotou_BP}).
\par A number of physical models have been proposed to explain Type II bursts, including among others, viscous instabilities and interactions between the disk and the rotating NS magnetosphere (e.g. \citealp{Taam_Evo,Spruit_Type2Mod}). None of the proposed models can fully reproduce the observed phenomenology \citep{Lewin_Bursts}, nor explain the significant differences observed between the BP and the RP (e.g. \citealp{Lewin_BP}, Court et al. \textit{in prep}). Most importantly, so far it is not understood what differentiates the RB and the BP from the more than 100 known NS-LMXBs (e.g. \citealp{Liu_Catalog}) which do not show Type II bursts. \citealp{DeMartino_XSS} previously noted similarities between Type II bursting and X-ray variability in the lightcurve of the TMSP 1FGL J1227.9-4852 \citep{Hill_XSS}. However, the authors note that the energetics are inconsistent with these phenomena being physically the same.
\par In Court et al. (\textit{in prep.}), we performed a detailed analysis of all archival X-ray data (including \textit{RXTE}, \textit{Swift}, \textit{Chandra}, \textit{XMM-Newton}, \textit{Suzaku}, \textit{NuStar}, \textit{Fermi} and \textit{INTEGRAL}) and found that the Type II phenomenology in the BP is much richer than previously thought (e.g. \citealp{Giles_BP}): the characteristics of the flaring evolve with time and source luminosity. Near the end of this evolution, we observed periods of highly--sctructured and complex high--amplitude X-ray variability. We refer to this variability as `Structured Bursting', which is unlike what is seen other LMXBs but very similar to the `hiccup' accretion observed in TMSPs.

\vspace{-2em} 

\section{Comparison}

\par \citealp{Rappaport_BPHistory} have previously suggested that the BP represents a slow X-ray pulsar nearing the end of its accreting phase. As such it is natural to compare this system with TMSPs, which are also believed to be systems approaching this evolutionary stage. In addition to this, \citealp{Degenaar_174457} have previously noted that the BP shows extended low-luminosity states during outburst, similar to those seen in the TMSP candidate XMM J174457-2850.3.

\begin{figure*}
 \centering
 \resizebox{1.80\columnwidth}{!}{\rotatebox{0}{\includegraphics[clip]{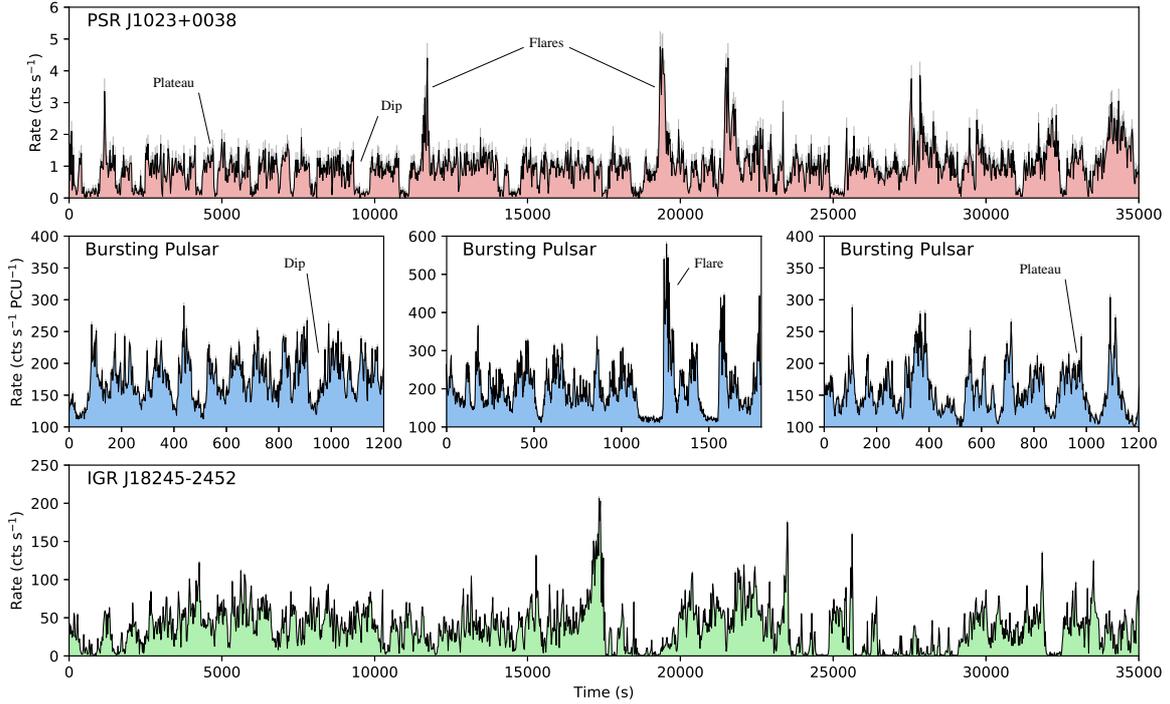}}}
 \caption{\small \textbf{Top:} 2--15\,keV \textit{XMM} lightcurve from the TMSP PSR J1023+0038. \textbf{Middle:} 2--60\,keV \textit{RXTE} lightcurves from the BP during its 1996 and 1997 outbursts, showing similar variability patterns to those seen in PSR J1023+0038. \textbf{Bottom:} 2--15\,keV \textit{XMM} lightcurve from the TMSP IGR J18245-2452. \textit{XMM} lightcurves are shown from 2--15\,keV so that they can be more directly compared with \textit{RXTE}.}
 \label{fig:lcs}
\end{figure*}

\par In Figure \ref{fig:lcs}, we show \textit{RXTE} lightcurves of `Structured Bursting' from the BP alongside lightcurves from periods of `hiccup' variability observed in the confirmed TMSPs PSR J1023+0038 and IGR J18245--2452. All three sources show similar patterns of X-ray variability: 
(i) \textit{Plateaus}: periods of approximately constant count rate with high-amplitude flicker noise (all plateaus in a given observation have approximately the same mean rate),
(ii) \textit{Dips}: Periods of low count rate ($\lesssim0.5$ of the rate in plateaus) with significantly less flicker noise, and 
(iii) \textit{Flares}: Relatively short-lived increases of the count rate to values $\gtrsim2$ times greater than the rate during plateaus.
In TMSPs, these features are interpreted as representing three quasi-stable accretion modes: the `high', `low' and `flaring' modes respectively (e.g. \citealp{Bogdanov_TMSPVar}). The most significant difference is that, in general, the variability in the BP occurs on timescales $\sim1$ order of magnitude longer than those in TMSPs.

%
%
%

\par
\begin{figure}
 \centering
 \includegraphics[width=.82\linewidth, trim={1.3cm 0.1cm 1.7cm 1.1cm},clip]{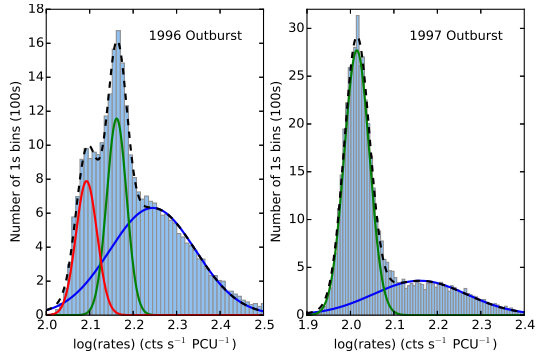}
 \caption{\small Histograms of the 1\,s binned count rates from all \textit{RXTE} observations of Structured Bursting in the 1996 (left) and 1997(right) outbursts of the BP. For the 1996 outburst, we fit the distribution with three Gaussians, while for the 1997 outburst we fit the distribution with 2 Gaussians. The individual Gaussians are plotted in solid lines, while the combined total is plotted in a dashed line.}
 \label{fig:bimodal}
\end{figure}

 In Figure \ref{fig:bimodal} we show histograms of the 1\,s-binned count-rate from all \textit{RXTE} observations of Structured Bursting in the 1996 (left) and 1997 (right) outbursts of the BP. As is the case for TMSPs, the histograms can be described with a number of log-Normally distributed populations: 3 populations in the 1996 outburst and 2 in the 1997 outburst. It is unclear why a population would be absent from the 1997 outburst, but some TMSPs have been observed to miss the `high' mode during hiccup accretion (e.g. IGR J18245-2452, \citealp{Ferrigno_TMSPVar}).

\par Detailed works on the low and high modes observed in the light curves of TMSPs show that X-ray pulsations are seen during both modes. Pulsations are fractionally weaker in the low state than the high state (for example varing between $4.0\pm0.2\%$ and $16.8\pm0.2\%$ in the TMSP IGR J18245-2452, \citealp{Ferrigno_TMSPVar}). In the case of the BP, we detect pulsations both during the low and the high modes; much like in TMSPs, the pulsations are weaker in the low mode. For example in \textit{RXTE} OBSID 10401-01-59-00 (in 1996), the pulsations had amplitudes of $3.5\pm0.2\%$ and $4.9\pm0.2\%$ respectively, while in OBSID 20078-01-23-00 (in 1997), the pulsations had amplitudes of $4.5\pm0.1\%$ and $6.0\pm0.1\%$ respectively. A reduction in pulse fraction in accreting pulsars has been interpreted as a change in accretion geometry due to a sudden decrease in the amount of matter reaching the compact object (e.g. \citealp{Ibragimov_PulseFrac}), and as such this result provides direct evidence that the Structured Bursting in the BP is caused by switches between accretion and propeller-driven outflows.

\par TMSPs are amongst the only LMXBs which are also significant $\gamma$-ray sources (e.g. \citealp{Hill_XSS}). The \textit{Fermi} point source 3FGL J1746.3--2851c is spatially coincident with the BP. While the field is too crowded to unambiguously associate 3FGL J1746.3--2851c with the BP, the existence of a $\gamma$-ray point source at this location is consistent with the possibility that the BP and TMSPs show the same phenomenology.

\par
\begin{figure}
 \centering
 \includegraphics[width=.82\linewidth, trim={0.6cm 0.1cm 1.0cm 1.1cm},clip]{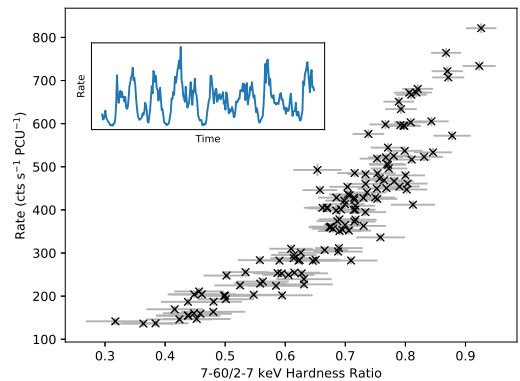}
 \caption{\small A 7--60/2--7\,keV hardness-intensity diagram for \textit{RXTE} observation 10401-01-59-00; the lightcurve of this observation is shown in the inset. To correct for the high background of the region, we subtract the median count rate of \textit{RXTE} observation 30075-01-24-00 from each band; at this time, GRO J1744-28 was in quiescence. We find a strong correlation between hardness and count rate, with a Spearman Rank Correlation Coefficient of 0.93. Data for the hardness-intensity diagram are binned to 10\,s, while data for the lightcurve are binned to 5\,s.}
 \label{fig:HR}
\end{figure}

 The spectral evolution of known TMSPs is varied. In PSR J1023+0038, the low, high and flaring modes all present similar spectra \citep{Bogdanov_TMSPVar}. However in IGR J18245-2452, \citealp{Ferrigno_TMSPVar} have found a strong correlation between spectral hardness and intensity during hiccups, showing that there is spectral evolution over time in this source. In Figure \ref{fig:HR} we show the hardness-intensity diagram of the BP during periods of Structured Bursting. We find a significant correlation, similar to what is seen in IGR J18245-2452 \citep{Ferrigno_TMSPVar}. This is in contrast with other slow accreting pulsar systems such as Vela X-1, which show an anticorrelation between these parameters during periods of variability \citep{Kreykenbohm_Vela}.



\vspace{-2em} 

%
%
%
%

\section{Discussion}

\par In this letter we compare the lightcurve, spectral and timing properties of the Bursting Pulsar at the end of its 1996 and 1997 outburst with those observed from Transitional Millisecond Pulsars. The data suggest that the BP may have undergone ``hiccup'' accretion similar to that seen in TMSPs, during which transferred matter alternates between being accreted onto the poles of the NS and being ejected from the system by the `propeller' effect (e.g. \citealp{Ferrigno_TMSPVar}). This similarity raises the exciting prospect of studying the physics of TMSPs in a completely different regime.

%
%

\par
\begin{figure}
 \centering
 \includegraphics[width=.82\linewidth, trim={0.6cm 0.1cm 1.0cm 1.1cm},clip]{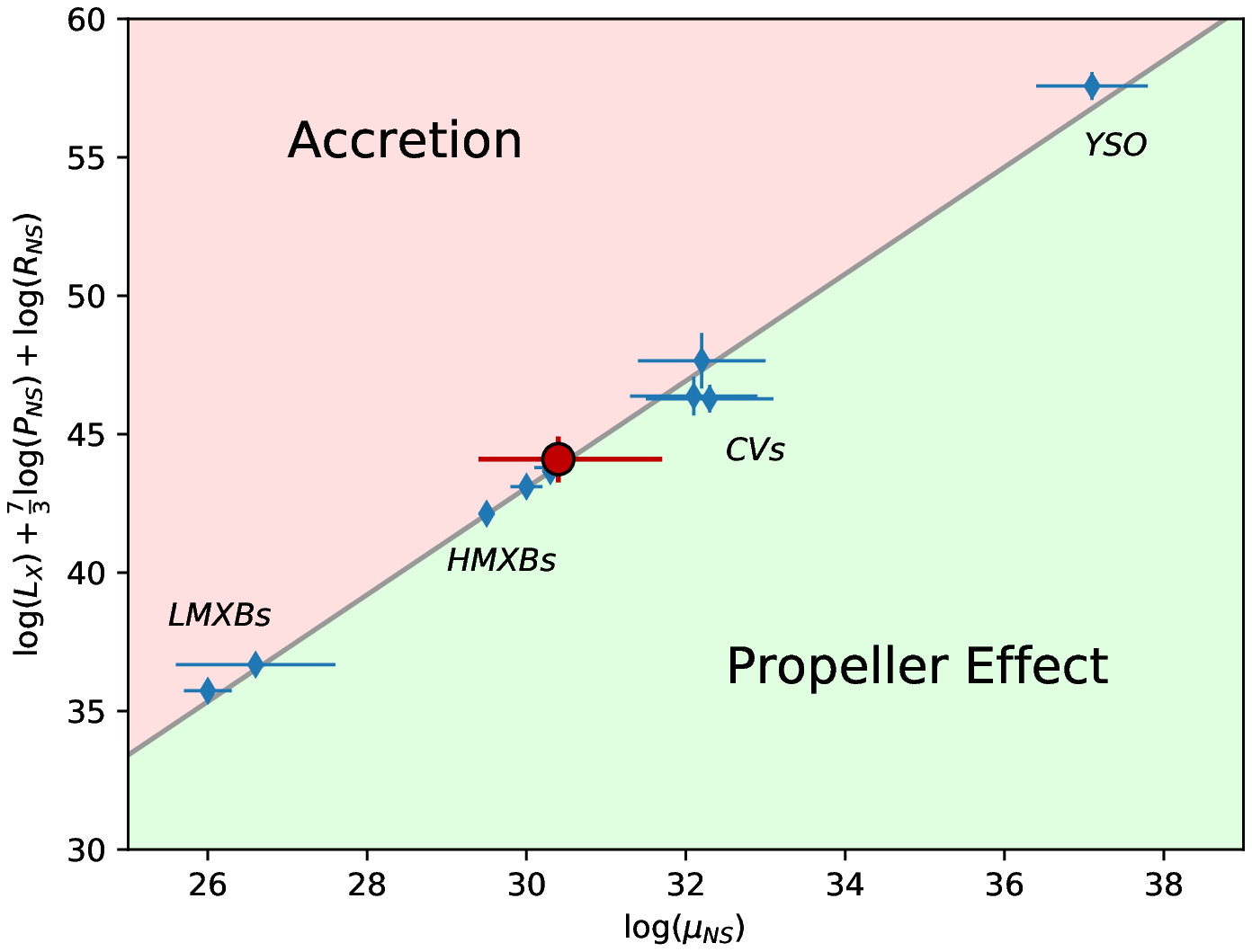}
 \caption{\small A plot of a number of objects ranging in scale from LMXBs and High-Mass X-ray Binaries (HMXBs) to Cataclysmic Variables (CVs) and Young Stellar Objects (YSOs) (blue diamonds). In each case, the object is plotted at the luminosity which defines its transition between propeller-mode accretion and free accretion. \citealp{Campana_PropBorder} suggest that any object above the line of best fit accretes freely, whereas all objects below are in the propellor regime. The BP (red circle) is consistent with approaching this line during periods of Structured Bursting. Errorbars on the BP represent the range of the reported magnetic fields as well as a range of stellar radii between 10--20\,km. The range in luminosity for the BP is calculated using 1.5-25\,keV \textit{RXTE}/PCA flux, assuming a distance of between 4--8\,kpc (e.g. \citealp{Kouveliotou_BP,Gosling_BPCompanion,Sanna_BP}) and a bolometric correction factor of 1--3.  Data on the other objects taken from \citealp{Campana_PropBorder}. $L$ is the bolometric luminosity of the object in ergs\,s$^{-1}$, $P$ is the period in s, $R$ is the radius in cm and $\mu$ is the magnetic moment in $Gauss\,cm^3$.}
 \label{fig:propBorder}
\end{figure}

 Very recently \citealp{Campana_PropBorder} proposed a universal relation between magnetic moment, spin frequency, stellar radius and luminosity at the boundary between accretion and the propeller effect. Any object that exists on one side of this boundary should be able to accrete, whereas objects on the other side should be in the propeller phase or not accreting at all. In Figure \ref{fig:propBorder} we reproduce \citealp{Campana_PropBorder}'s results and include our estimates for the BP during the periods of Structured Bursting. We find that the BP is consistent with lying on or near the boundary between propeller-mode and direct accretion, clustering with High Mass X-ray Binaries (as expected due to the BP's high magnetic field), and supporting the link between ``hiccups'' and Structured Bursting.

\par If the ``hiccups'' in the BP show that the system is transiting to a radio pulsar, then the BP should not lie in the $P$-$\dot{P}$ `graveyard' region \citep[e.g.][]{vandenHeuvel_Graveyard}. To our knowledge, there is no measurment yet of the NS spin down during the BP's X-ray quiescent state. Under the assumption that the BP becomes a radio pulsar, and that the possible spin down during that period is due to the same mechanism as those of the known radio pulsars, we can position the BP in the $P$-$\dot{P}$ diagram (the plot of pulsar spin $P$ against spin-down rate $\dot{P}$, not shown) by using the orbital period and estimates of its magnetic field. At $B\sim2\times10^{11}$G, the BP falls well outside of the pulsar graveyard. We note that \citet{Pandey-Pommier_BPRad} and \citet{Russell_BPRad} did not detect a significant radio source at the location of the BP during X-ray outburst. To our knowledge, there is no report of Radio detection/non-detection during X-ray quiescence.


%
%

\par In addition to the BP, several additional sub-10\,Hz accreting X-ray pulsars have been discovered (e.g. GX 1+4 and 4U 1626-67, \citealp{Lewin_GX1,Rappaport_4U}). The reason behind the slow spins of these objects is poorly understood, but a number of these systems have been seen to undergo `torque reversal' events, during which $\dot{P}$ switches sign (e.g. \citealp{Chakrabarty_4U,Chakrabarty_GX14}). In some sources, the magnitude of the spin-down during an event is of the same order magnitude as the preceding period of spin-up, resulting in little or no net spin change. Torque reversal events occur irregularly, but the recurrence timescale varies between objects from weeks to decades (e.g. \citealp{Bildsten_Rev}).


%
%

\par Given that the BP has a strongly stripped stellar companion \citep{Bildsten_Nuclear}, a high magnetic field and shows significant spin-up during outburst (e.g. \citealp{Finger_BP,Sanna_BP}), it is difficult to explain its low spin by suggesting the system is young or that the angular momentum transfer is inefficient. \citealp{Rappaport_BPHistory} suggest that the magnetic field and spin could be explained if much of the mass transfer in the system occurred before the primary became a neutron star, but they note that this scenario is inconsistent with the low mass of the donor star.
\par Torque reversal events in the BP (similar to those seen in other slow accreting pulsars, e.g. \citealp{Bildsten_Rev}) could explain why the pulsar has failed to reach a spin rate on par with TMSPs. Although no torque reversal event has been reported from the BP, it is feasible that the recurrence timescale of such an event is longer than the $\sim20$ years for which the object has been studied (this is consistent with the recurrence timescales seen in other slow accreting pulsars). The discovery of torque reversal in the BP would strongly link it with the other known slow accreting pulsars. 

%
%

%
%


\par The BP has a spin rate $\sim2$ orders of magnitude less than previously known TMSPs, and a magnetic field $\sim2$ orders of magnitude stronger, but it still shows lightcurve, timing and spectral behaviour which are remarkably similar to TMSPs. This raises the exciting prospect of exploring the physics of TMSPs in a previously unexplored physical regime. If the BP itself is a transitional pulsar, it should emit radio pulsations during X-ray quiescence. Future detections of radio pulsations from this object would unambiguously confirm it as a transitional pulsar.

\vspace{-2em} 
\bibliographystyle{mnras}
\bibliography{/home/jamie/Documents/Bibliographies/refs}





\bsp	
\label{lastpage}
\end{document}